# A COMPARISON OF THE KNOWLEDGE-BASED INNOVATION SYSTEMS IN THE ECONOMIES OF SOUTH KOREA AND THE NETHERLANDS USING TRIPLE HELIX INDICATORS[1]

Han Woo Park[2]
*YeungNam University*
*South Korea*

Heung Deug Hong[3]
*KangWon National University*
*South Korea*

Loet Leydesdorff[4]
*University of Amsterdam*
*the Netherlands*

*Abstract*

This paper elaborates on the Triple Helix model for measuring the emergence of a knowledge base of socio-economic systems. The 'knowledge infrastructure' is measured using multiple indicators: webometric, scientometric, and technometric. The paper employs this triangulation strategy to examine the current state of the innovation systems of South Korea and the Netherlands. These indicators are thereafter used for the evaluation of the systemness in configurations of university-industry-government relations. South Korea is becoming somewhat stronger than the Netherlands in terms of scientific and technological outputs and in terms of the knowledge-based dynamics; South Korea's portfolio is more traditional than that of the Netherlands. For example, research and patenting in the biomedical sector is underdeveloped. In terms of the Internet-economy, the Netherlands seem oriented towards global trends more than South Korea; this may be due to the high component of services in the Dutch economy.

---

[1] The authors acknowledge that this research was supported by the Korean Research Foundation (KRF-2003-042-H00003). They are grateful to Sung Jo Hong for his collaboration in the project.
[2] hanpark@yumail.ac.kr
http://www.hanpark.net
[3] hdhong@kangwon.ac.kr
[4] loet@leydesdorff.net
http://www.leydesdorff.net

## INTRODUCTION

As new network technologies such as the Internet have permeated society, they become another driving force changing the form of the economy of a nation. New technologies enable individual and institutional actors to collaborate in additional modes, but these processes make them increasingly interdependent in terms of the information exchange. New patterns of collaboration with potential competitive advantages can then be developed. Gibbons *et al.* (1994) have called this type of knowledge organization and production "Mode 2". An overlay of communications and knowledge-based expectations is increasingly added to the existing institutions.

While the political economy coordinated two functions—notably the market and the state—the knowledge infrastructure coordinates the three subdynamics of (i) wealth production, (ii) organized novelty production, and (iii) private appropriation versus public control. In other words, political economies are increasingly transformed into knowledge-based economies by the additional subdynamics of systematically organized innovation processes. Because national economies are open systems surrounded by an external environment, they interact with a variety of elements in the society. The resulting dynamics are complex (at the phenotypical level) and can therefore no longer be expected to contain central coordination.

The *knowledge infrastructure* of national innovation systems can be operationalized in terms of networks. The network approach can be used for identifying the structures in social systems based on the relations among the system's components rather than the attributes of individual cases (Latour, 1987; Wasserman & Faust, 1994). This approach can be generalized to describe the structures of the knowledge-based innovation systems in the national economies. For example, Etzkowitz & Leydesdorff (1997) proposed to consider the networked knowledge infrastructure in terms of a Triple Helix of university-industry-government relations. The networks provide us with a fingerprint of the innovation system at each moment in time. They contain the expected information of the evolving *knowledge base* that can be conceptualized as the result of the fluxes of communication and information constrained and enabled by these networks.

Historically, the advanced industrial nations first generated 'national systems of innovation' during the period 1870-1980 (Freeman, 1988; Lundvall, 1988, 1992; Nelson, 1993). The innovative knowledge flows within these political economies, however, span boundaries and thus generate new types of competition at the global level (Krugman, 1996; Leydesdorff, 2001; Sahal, 1981; Schumpeter, [1939] 1964). In the Triple Helix model this selection pressure is represented as a networked overlay of communications among the institutional agencies which have hitherto carried the knowledge infrastructure along trajectories: industry, academia, and government. Each of these institutions is organized along international dimensions as well. The overlay of expectations, however, functions as a next-order regime level (Dosi, 1982). At this level, the institutional participants can entertain and recombine possibilities other than those that have been realized hitherto. Nations can then be considered as niches competing in the international arena in terms of their innovative capacities (Etzkowitz & Leydesdorff, 2000; Leydesdorff & Etzkowitz, 2003).

In other words, we are living in what can be considered a 'post-industrial' society because the system is no longer local, but knowledge-based and hence continuously globalizing (Beniger, 1986; Bell, 1973; Giddens, 1990; Toffler, 1980). In the post-industrial society a plethora of information and knowledge has continuously to be managed. One witnesses the explosion of information and knowledge produced and distributed by the traditional forms of knowledge supplier systems such as universities as well as research and development sectors in both public and private organizations, government institutions, and pressure groups. In this configuration, it is essential for a nation that the institutional retention mechanism is adapted to the evolving knowledge base (Freeman & Perez, 1988). How can the vast amount of available knowledge be gathered, generated, and enriched by the network of organizations involved so that the knowledge can be applied in many different and varied circumstances. Under which conditions can networking strengthen a national system of innovations? (Leydesdorff, 2002).

Knowledge has been necessary in the functioning of any society. However, the organization of the production of knowledge at the social level (e.g., in R&D laboratories) has been typical for the industrial society (Whitley, 1984). The post-industrial society produced the sophisticated digital technologies such as the Internet which have hastened our plunge into the knowledge-based society. New technologies affect knowledge creation and diffusion in a number of ways, which lets society as a whole shift to knowledge-intensive activities. Using improved computing technologies and digital networks, knowledge-based activities can be performed in cooperation with other components in social systems in almost an infinite number of ways when and how they are needed (Shapiro & Varian, 1999; Steinmueller, 2002). Information can be codified into knowledge and then also be commodified. Thus, it might be appropriate to say that we live in a digital knowledge-based society.

## NATIONAL SYSTEMS OF INNOVATION

The Triple Helix model enables us to study the network linkages among industry, academia, and government both in the evolutionary terms of the transition to post-industrialism and in terms of communication-theoretical concepts (Leydesdorff, 2001). In this study, we apply recent advances at the methodological level for studying this complex arrangement to two national systems of innovation, namely South Korea and the Netherlands as emerging knowledge-based economies. Both these national systems are highly innovative, but they are also sufficiently different so that we are able to explore how our operationalization into indicators performs in different contexts. More specifically, we will apply visualization techniques that we developed in other contexts to the first-order indicators like patents and scientific publications and we will use the data gathered in this way for the second-order evaluation in terms of Triple-Helix relations.

As noted, the transition to a knowledge-based economy requires the transformation of the political economy. While the latter is mainly based on the coordination between private capital and public control, the systematic organization of innovation at the social level continuously upsets and transforms the public/private interfaces in new arrangements among heterogeneous partners. The function of government itself then also has to change. The continuation of South Korea's high economic growth, for example, will increasingly depend on technological innovations produced within South Korea. As a result, the Korean government is placing great emphasis on stimulating indigenous technical advances and on making the economy more conducive to innovation. In the meantime South Korea is as wired and digitalized as the North American and European nations that play more prominent roles in the knowledge-based economy, but the production, use, and application of knowledge is still lower than in these other countries.

The Netherlands is part of the European Union. This supra-national level of government provides an additional coordination mechanism and incentive for organizing technological innovation and social transformation (Freeman & Perez, 1988; Laredo, 2003; Frenken & Leydesdorff, 2004). Furthermore, the Netherlands has been a center of trade and knowledge reproduction for centuries. Its industrial base is relatively weak in comparison to its European partners. The transformation into a knowledge-based economy has been a priority for both government and private industry (NOWT, 2000).

## MEASURING THE KNOWLEDGE-BASE OF INNOVATION SYSTEMS

The knowledge base of an economy is not a given state, but remains operational as a driver of change. During this evolutionary reconstruction elements from different sources are recombined under the pressure of economic competition. The network of university-industry-government relations can be considered as an institutional "knowledge infrastructure" that carries a system of operations containing science, technology, and knowledge-based innovations. These three domains (science, technology, and innovation) can as a first-order (institutional) approximation be measured using different indicators: technology indicators (e.g., patents), scientometric indicators, and communications at the Internet (webometric analysis). The information contents in these three dimensions can thereafter be recombined and thus enrich our understanding of the national system as a specific form of integration. Figure 1 summarizes this idea using a visual representation.

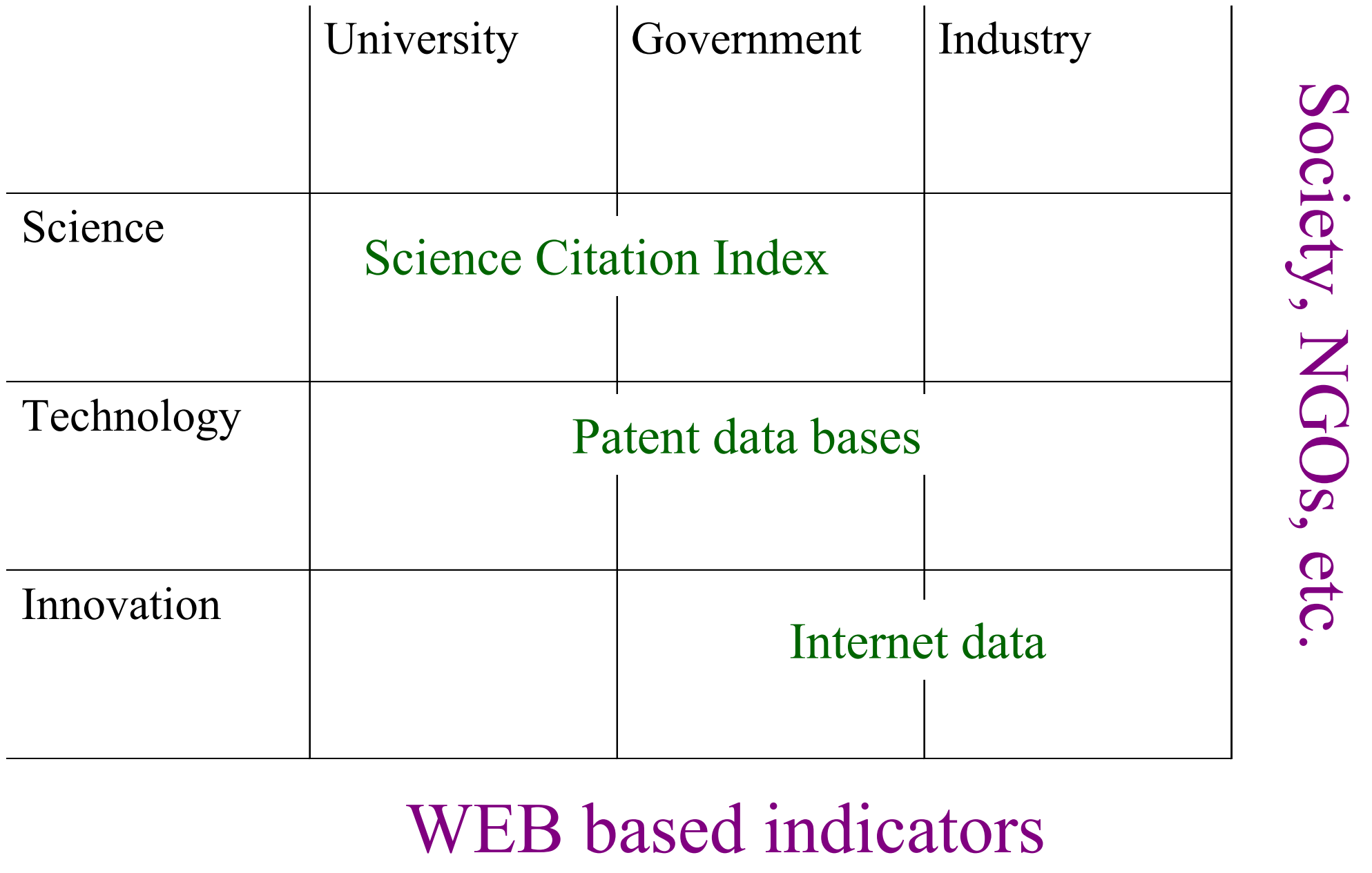

**Figure 1:** Institutional and functional differentiation in the Internet age (from: Leydesdorff & Scharnhorst, 2003)

Methodologically, the mutual information in the three dimensions of the Triple Helix provides us with a measure for networks at each moment in time in terms of probability distributions and to evaluate the measurement results in terms of the dynamics. The description of the network data in terms of probability distributions enables us to use Shannon's (1948) mathematical theory of communication. A probability distribution contains an uncertainty. The expected information content of the message that these events have happened with this observed frequency distribution, can be expressed in terms of bits of information using the Shannon-formulas (Abramson, 1963; Theil, 1972; Leydesdorff, 1995). The mutual information between two dimensions of the probability distribution (for example, in university-industry (UI) relations) is then equal to the transmission (T) of the uncertainty (Theil, 1972):

$$T_{UI} = H_U + H_I - H_{UI}$$

The relationship reduces the uncertainty for the two relating systems (with $H_{UI}$). Abramson (1963, at p. 129) showed that the mutual information in three dimensions can be derived as:

$$T_{UIG} = H_U + H_I + H_G - H_{UI} - H_{IG} - H_{UG} + H_{UIG}$$

Note that the uncertainty of the variables measured in each of the interacting systems ($H_U$, $H_I$, and $H_G$) is reduced at the systems level by the relations at the interfaces between them, but the three-dimensional uncertainty adds positively to the uncertainty that prevails. Because of this alteration of the signs, the three-dimensional transmission can become negative. As noted, this reduction of the uncertainty by the negative transmission is exclusively a result of the network configuration of bi-lateral relations that develop without central coordination.

Triple Helix configurations can first be depicted *statically* using social network analysis or in more general terms, as semantic network diagrams. Social network analysis is particularly useful for discovering hidden patterns that could not be found if research objects (often called nodes) are analyzed individually (Wasserman & Faust, 1994). In this study, we apply social network visualization techniques in the analysis of title words of scientific articles, patents, and their literature references. The analysis will be done using the algorithm of Kamada and Kawai (1989) as it is available in the software package Pajek. This algorithm represents the network as a system of springs with a relaxed length proportional to the edge length. Nodes are iteratively repositioned to minimize the overall "energy" of the spring system using a steepest descent procedure. In order to keep the visualizations readable, the analysis will pragmatically be limited to the approximately one hundred most frequently occurring words for each case. Unconnected nodes are therefore not included in the visualizations below. Note that network maps can be much more finegrained than Triple-Helix statistics because of the much higher level of visualization for some relationship among the three dimensions of the Triple Helix.

The outline of this paper is that the representation of a Triple Helix *dynamics* based on the *AltaVista* search results at the Internet is first introduced. Thereafter, we turn to science indicator using the *Science Citation Index* to measure Triple-Helix status of South Korea and the Netherlands with reference to international levels. The Triple Helix measures are developed both with reference to science indicators and with reference to *AltaVista* data. Finally, the technological levels of the two countries are measured using patent data. Additionally, the mapping of title words in terms of social network analysis can be compared between the patent and the publication data.

## RESULTS

### a. Innovation Indicators

*Data-gathering methods*

The degree of innovation in the knowledge base of an economy can also be measured using a webometric approach (Park & Thelwall, 2003). Similar to scientometric analysis (Garfield, 1979), the webometric approach quantitatively evaluates the scale of the Web in terms of co-words. Although the Web can be considered as a globalized system, it can be searched specifically for national domains using the ccTLD (country code Top Level Domain), for example, using ".kr" for South Korea and/or the national language, that is in this case, Korean. Following the scheme of Leydesdorff & Curran (2000) and Leydesdorff (2003), we explored the various dimensions of university-industry-government relations using the AltaVista Advanced Search Enginge. In this study, the two national domains with their respective languages are analyzed: South Korea (.kr) and the Netherlands (.nl) with Korean and Dutch as the respective languages.

All searches were conducted on April 26, 2004.[5] More specifically, we made a search specific to the national domains of South Korea and the Netherlands with words in the vernacular language meaning university, industry, and government, respectively. When using Dutch as the search language, we used as search terms: "universiteit," "industrie," and "overheid." The literal translation of "government" into Dutch is "regering," but relations with a "regering" would mean that the relations are maintained with a specific (that is, politically elected) administration. The institution of government is more precisely expressed by the Dutch word "overheid."

For South Korea, we used "대학(dae-hwag)," "기업(ghi-oeup)," and "정부(jeong-bu)" as search terms. The literal translation of "industry" into Korean is "산업(san-oeup)," but "기업(ghi-oeup)" would be more suitable term in the context of Triple Helix relations after consultation with a number of native speakers. Additionally, the combination of gTLDs (generic Top Level Domains) provided us with a system of reference for assessing the relative contributions of these two nations globally. In other words, a total of the 14 generic extensions (.com, .net, .org, .edu, .gov, .int, .mil, .biz, .info, .name, .pro, .aero, .coop, and .museum) were combined into an international system of reference by using Boolean OR-operators. English was used as the search language for the international domain.

*Results*

Even with a limited amount of data the number of possible comparisons and analyses is large. For example, one can compare among the countries, over time, in terms of using different languages, and in terms of bilateral and trilateral relations, using the various options of the search engine.

Let us first explore the differences between South Korea, the Netherlands, and the combined gTLDs in terms of the number of hits for the three Triple-Helix categories. The respective sizes of the documents sets using these three words (university, industry, and government) as retrieval terms are considerably different among the three domains (Table 1).

[5] *AltaVista* uses the *Yahoo!* search engine since April 2004. Our searches are from after this transition.

First, almost all sets are more than twice as large in the case of South Korea compared with the Dutch contributions. The words “university” and “industry” are much more dominant in South Korea than in the Netherlands. The differences between these countries were the smallest in the word “government.” However, the roles of these words in the South Korean domain are not greatly visible with reference to the combined gTLDs. Overall, the word “university” is the leading keyword across domains. This may be due to the fact that the early use of the Internet was predominantly academic and the majority of Internet users are nowadays university students.

**Table 1**: Number of hits for Triple Helix components in three domains

| Year | Domain-Language | University | Industry | Government |
|---|---|---|---|---|
| 1999 | kr-Korean | 10582 | 6103 | 4947 |
| | nl-Dutch | 2931 | 1282 | 3186 |
| | gTLDs-English | 111197 | 45375 | 60903 |
| 2000 | kr-Korean | 19552 | 12956 | 9976 |
| | nl-Dutch | 6467 | 2700 | 6186 |
| | gTLDs-English | 168887 | 79827 | 98625 |
| 2001 | kr-Korean | 38191 | 29364 | 24925 |
| | nl-Dutch | 9534 | 5086 | 10236 |
| | gTLDs-English | 278909 | 156790 | 196442 |
| 2002 | kr-Korean | 45368 | 31021 | 17056 |
| | nl-Dutch | 17215 | 9558 | 17618 |
| | gTLDs-English | 489470 | 320542 | 366043 |
| 2003 | kr-Korean | 81535 | 74567 | 33958 |
| | nl-Dutch | 35523 | 22213 | 39594 |
| | gTLDs-English | 1281321 | 904392 | 1121088 |

*Searched with the *AltaVista Advanced Search Engine* on April 26, 2004

Table 2 provides the comparisons among the combinations of the three words “university,” “industry,” and “government” with Boolean AND-operators. The correspondence between the number of hits for South Korea and for the Netherlands reveals that the relations among individual Triple Helix components is more salient in the former rather than in the latter country. However, it is very clear that the combinations among “university,” “industry,” and “government” in South Korea lag a lot behind when compared to the global data set. Note that the pattern is almost identical for the counts of individual Triple Helix components as examined above. While the combination “university AND government” remains the most frequently occurring term in South Korea, the grouping of “university AND industry” was most prominent in the case of the Netherlands. The “university” and “industry” dyad was also the most important term in the reference domain of gTLDs.

**Table 2:** Number of hits for Triple Helix combinations in three domains

| Year | Domain-Language | U-I-G | Univ-Ind | Univ-Gov | Ind-Gov |
|---|---|---|---|---|---|
| 1999 | kr-Korean | 437 | 1129 | 1089 | 1391 |
| | nl-Dutch | 73 | 165 | 373 | 277 |
| | gTLDs-English | 7776 | 15081 | 21406 | 16012 |
| 2000 | kr-Korean | 624 | 2117 | 1783 | 2408 |
| | nl-Dutch | 110 | 320 | 876 | 529 |
| | gTLDs-English | 11550 | 22755 | 32707 | 25433 |
| 2001 | kr-Korean | 1210 | 3979 | 3404 | 4936 |
| | nl-Dutch | 197 | 508 | 1165 | 1084 |
| | gTLDs-English | 19115 | 38268 | 57838 | 46869 |
| 2002 | kr-Korean | 1444 | 5157 | 3750 | 5245 |
| | nl-Dutch | 398 | 945 | 2061 | 1792 |
| | gTLDs-English | 32084 | 66044 | 98211 | 89881 |
| 2003 | kr-Korean | 3316 | 9637 | 7382 | 11609 |
| | nl-Dutch | 860 | 2105 | 4213 | 3912 |
| | gTLDs-English | 75812 | 157968 | 235252 | 217334 |

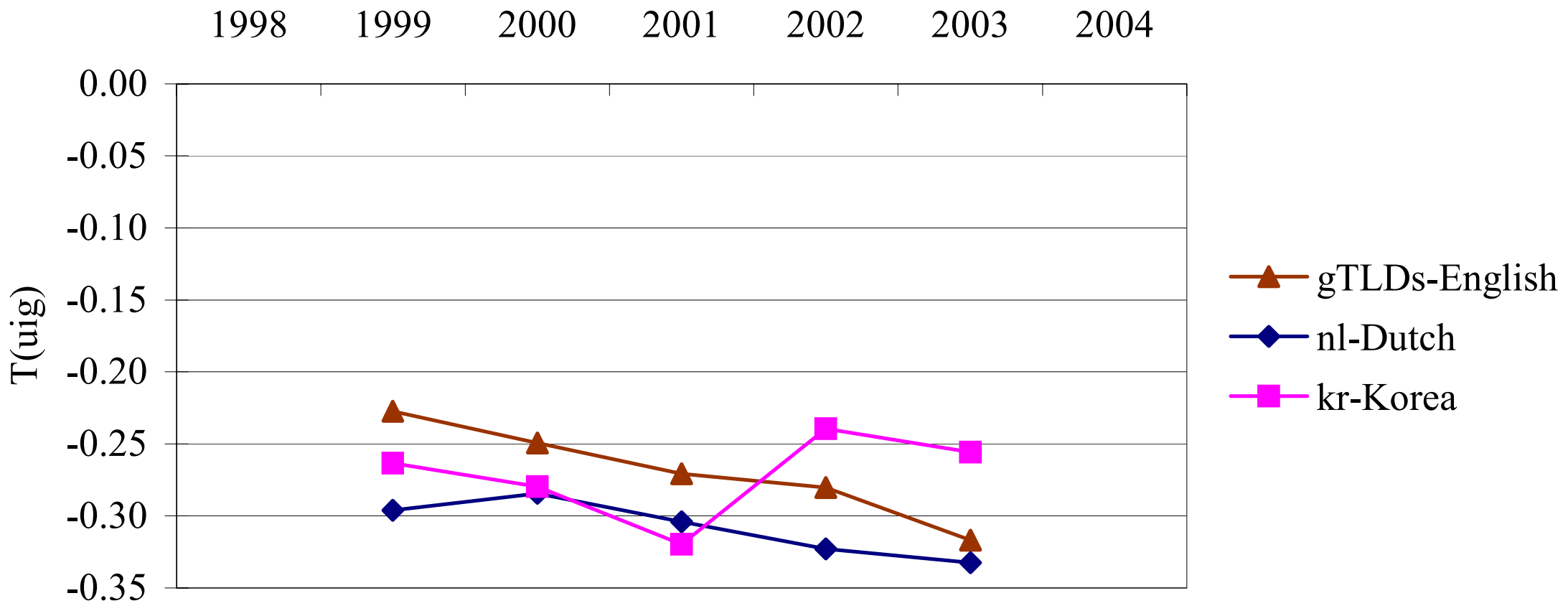

**Figure 2**: Three-dimensional transmission for South Korea and the Netherlands, respectively. (Two-year moving averages added.)

One can compute a three-dimensional transmission of Triple Helix relations for the various national systems and the respective languages during the period 1999-2003. As shown in Figure 2, the results of this calculation are striking. The Triple Helix overlay operated within the Netherlands and South Korea at a similar level of self-organization until 2001. In 2001, however, one can observe a discontinuity in the South-Korean curve. This may be caused by the collapse of dot com bubble in South Korea. Figure 3 shows that this was the case, indeed. Thus, the indicator flagged a substantial difference in the underlying dynamics.

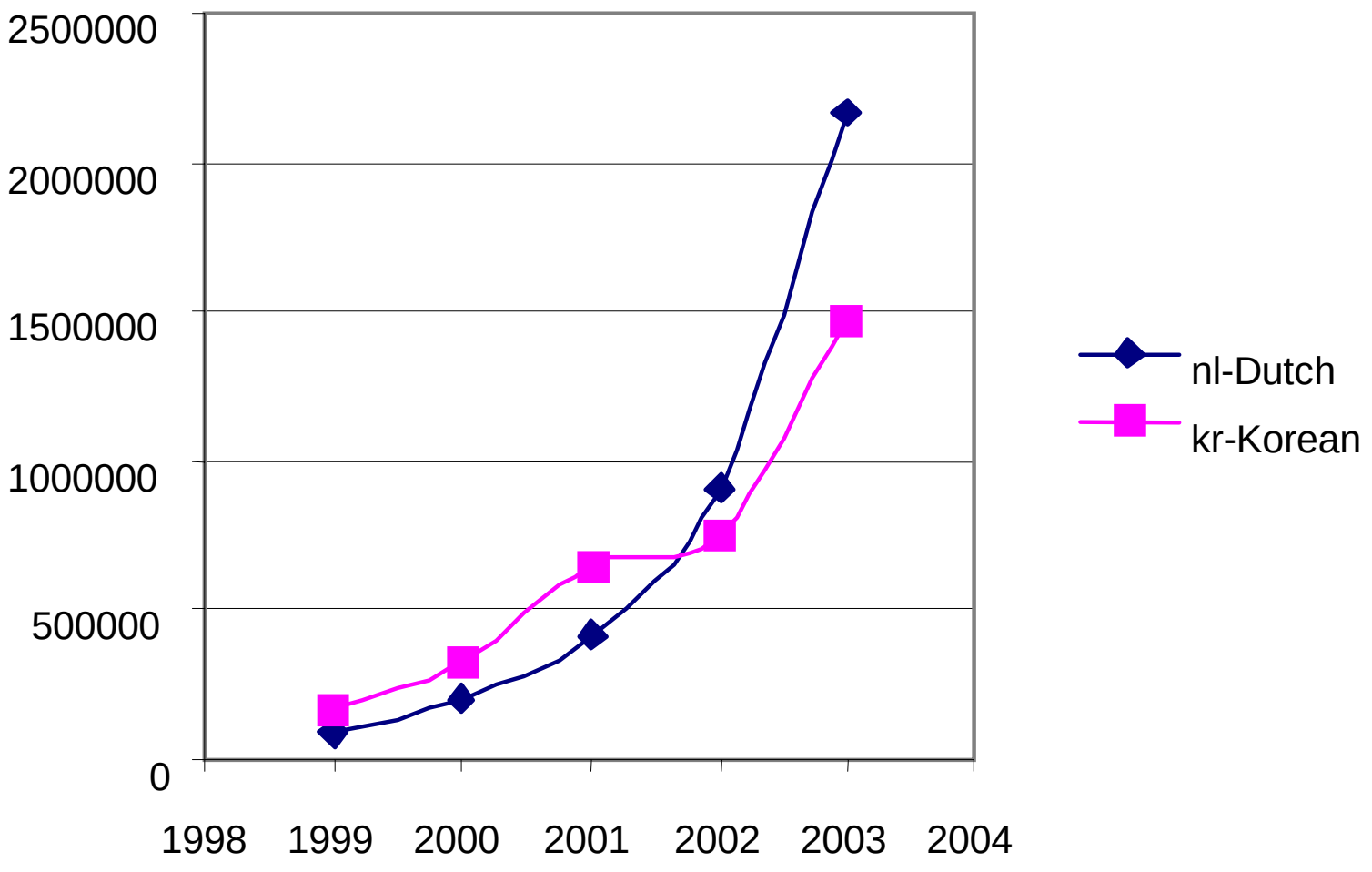

**Figure 3**: Total number of hits using Dutch or Korean as national languages at the Internet. (*AltaVista Advanced Search Engine*, 26 April 2004.)

In summary, South Korea has been less integrated from the perspective of Triple Helix. However, Figures 2 and 3 show that South Korea returned to a normal pattern in 2003. In terms of the three-dimensional transmission, the Netherlands is more networked in a Triple Helix mode than South Korea. The relations between university, industry, and government in the Netherlands are very similar to gTLDs domains. The time series for the Netherlands is almost identical to the global data set. However, it should be noted that the figures cannot be compared directly because the search terms are semantically different using the various languages. The semantic similarity between the English search terms in the gTLDs and those in the Netherlands may add a linguistic effect, possibly disadvantaging Korean language.

**b. Science Indicators**

*Data-gathering methods*

The scientific literature was the first system to be made subject to bibliometric analysis (Garfield, 1979; Leydesdorff, 1995; Narin, 1976). The *Science Citation Index* and its counterparts in the social sciences and the arts & humanities have become the standard for scientometric analyses to such an extent that funding decisions are often influenced by the status of research groups in these databases (Van Raan, 1988). In addition to ranking authors and institutions in terms of numbers of publications and/or citations, the databases also enable us to "map the sciences" using co-authorship relations, co-citations, co-words, etc. These mappings can be done in terms of institutional units (research groups, institutes, countries) using the address fields or in terms of cognitive units, for example, using the aggregated citation relations among scientific journals.

Following up on a study of Leydesdorff (2003) who used the *Science Citation Index* 2000 for computing the mutual information in three dimensions, we used corporate addresses on the CD-Rom version of the *Science Citation Index* 2002. Here, our research focuses on University-Industry-Government relations in this data set. An attempt was made to organize all these addresses automatically in terms of their attribution to university-industry-government relations. The routine first attributed a university label to addresses that contained the abbreviations .UNIV. or .COLL. Once an attribution was made, the record was set aside before further attributions were made. The remaining addresses were subsequently labeled as industrial if they contained one of the following identifiers CORP, INC, LTD, SA or AG. Thereafter, the file was scanned for the identifiers of public research institutions using NATL, NACL, NAZL, GOVT, MINIST, ACAD, INST, NIH, HOSP, HOP, EUROPEAN, US, CNRS, CERN, INRA, and BUNDES as identifiers.[6] This relatively simple procedure enabled us to identify percentages of the SCI literature in terms of their origin as university, industry, or government. However, these results remain statistically approximate figures.

*Results*

Table 3 is based on the *Science Citation Index* 2002. For a longitudinal comparison, we added the analysis for 2000 to the table. While the Netherlands is relatively declining in 2002 (when compared with 2000), the number of papers in Korea is increasing in all the categories. South Korea has even surpassed the Netherlands in terms of the number of University-Industry coauthored papers.

According to the results exhibited in Table 3, the addresses in the 2002 database point to 14,931 records with South Korea as their country names and 17,865 items contain a Dutch address. This corresponds to 2.2% and 2.6% of all university papers in these two subsets, respectively. South Korea and the Netherlands exhibited some similarities in the proportional pattern of individual Triple Helix elements. The number of records with a university address is the largest among the three domains in both South Korea and the Netherlands. The number of government addresses followed. Less than 1000 of the records contain industry addresses.

For all these subsets a three-dimensional transmission of Triple Helix relations can be calculated. The results of this calculation are also shown in Table 3. Table 3 suggests a very different pattern for the Triple Helix developments from 2000 to 2002 in South Korea and the Netherlands. In terms of the three-dimensional transmission in 2000, South Korea is more networked in a Triple Helix mode than the Netherlands. The T(uig) indicators are - 40.1 for South Korea and - 25.4 for the Netherlands in 2000. In 2002, South Korea is still doing a bit better than the Netherlands in terms of Triple-Helix dynamics, but this dynamics has gained relative weight in the Netherlands. (Or perhaps, one should say that this dynamic is less sensitive to the decline than the institutional ones.) The three-dimensional T-values are - 33.7 for Korea and -32.8 for the Netherlands. Interestingly, the value for the Netherlands may have been improved because of the ongoing decline of university output. Thus, the other Triple Helix partners gain in relative weight and the relations become relatively more important. For example, the number of papers with university addresses in the Netherlands is 16,379 for 2000 against 15,927 for 2002.

[6] The attribution of "institute" to the public research sector is doubtful in some cases. For example, the Korea Advanced Institute of Science and Technology (KAIST) can be also considered as part of the academic system.

**Table 3:** Comparison between South Korea and the Netherlands in Science Citation Index

| Country | Year | Number | % titles retrieved | T(uig) in mbits | UI | UG | IG | UIG | Univ | Ind | Govt |
|---|---|---|---|---|---|---|---|---|---|---|---|
| All | 2000 | 676511 | 93.3 | -77.0 | 16270 | 108919 | 4359 | 5201 | 543123 | 41242 | 232096 |
| | 2002 | 683222 | 93.6 | -70.7 | 17095 | 116782 | 4626 | 5664 | 556370 | 41840 | 234843 |
| South Korea | 2000 | 12038 | 98.3 | -40.1 | 351 | 2341 | 87 | 91 | 10345 | 676 | 3978 |
| | 2002 | 14931 | 98.7 | -33.7 | 533 | 3115 | 118 | 183 | 13163 | 996 | 4904 |
| Netherlands | 2000 | 18357 | 95.3 | -25.4 | 372 | 4482 | 106 | 259 | 16379 | 863 | 6593 |
| | 2002 | 17865 | 95.1 | -32.8 | 328 | 4663 | 78 | 307 | 15927 | 859 | 6762 |

In South Korea, the ratio of university papers coauthored with government was 20.9% compared to 19.4% at the measurement in 2000. In this case, the figures for the Netherlands went up from 20.4% to 26.1%. The coautorships among university, industry, and government rose in the Netherlands to 1.7% from 1.4% in 2000 and also the one in South Korea increased from 0.8% to 1.2%. But the interconnection between the industry and public sector research has become even stronger in South Korea over time (0.7% → 0.8%) contrary to a decrease in the Netherlands (0.6% → 0.4%). These data demonstrate how the participation of industry and the public sector in a university-driven knowledge production system can function as the crucial variable for the self-organization of the Triple Helix dynamics (Godin & Gingras, 2000).

It should be kept in mind that these results refer to representations in the *Science Citation Index*, and the above classifications into sectors were statistical and therefore approximate. For example, industry is weakly represented in this data. Collaborations of university researchers with industrial partners may often not lead to this type of scientific publication.

**Table 4.** Triple Helix index for various countries and regions in Science Citation Index

| *Year* | *2000* | *Rank* | *Year* | *2002* |
|---|---|---|---|---|
| *Countries* | *T(uig) in mbits* | | *Countries* | *T(uig) in mbits* |
| Japan | -92.1 | 1 | Japan | -82.4 |
| India | -78.1 | 2 | USA | -71.0 |
| USA | -74.4 | 3 | India | -67.7 |
| UK | -63.1 | 4 | UK | -54.0 |
| France | -52.1 | 5 | EU-15 | -45.3 |
| EU-15 | -50.1 | 6 | France | -42.5 |
| Germany | -43.4 | 7 | Germany | -39.6 |
| S Korea | -40.1 | 8 | S Korea | -33.7 |
| Scandinavia | -31.6 | 9 | Netherlands | -32.8 |
| Italy | -29.4 | 10 | Scandinavia | -32.5 |
| Netherlands | -25.4 | 11 | Singapore | -28.6 |
| Russia | -24.2 | 12 | Italy | -27.6 |
| Singapore | -23.9 | 13 | Brazil | -26.8 |
| Brazil | -22.4 | 14 | Russia | -18.9 |
| Taiwan | -17.1 | 15 | Taiwan | -18.0 |
| PR China | -14.9 | 16 | PR China | -11.0 |

Table 4 shows the global pattern of Triple-Helix development during the years 2000 and 2002. Among the 16 countries and regions listed, Japan has the highest value on the Triple-Helix index in both years. The U.S.A. and India follow as a salient group in the development of mutual relations among Triple-Helix components. South Korea occupied the 8th position in both the year 2000 and 2002 while the Netherlands jumped from 11th to 9th place. Despite the same ranks for South Korea in two years, the values of Triple-Helix development are quite different (-40.1 for the year 2000 versus –33.7 for the year 2002). The difference could be explained by the fact that the role of universities in South Korea was too dominant at the network level. Interestingly, the Triple Helix overlays within People's Republic of China operate at a much lower level of self-organization than in various world regions. Overall, the Triple-Helix measures in the year 2002 are quite similar to those of the year 2000.

In 2002, there were 15,127 Korean scientific publications with at least one Korean address among authors while there were 18,792 articles with a Dutch address. However, the number of word occurrences in the titles is little bit higher in Dutch publications than in Korean ones (177,707 versus 144,597 words). Table 5 provides the comparison of the Korean and Dutch data sets in the *Science Citation Index 2002* for the purpose of the semantic network mapping which we pursue below on the basis of this data.

**Table 5:** Comparison between South Korea and the Netherlands in unique words

| Items | *South Korean address* | *Dutch address* |
|---|---|---|
| Number of records in the SCI 2002[7] | 15127 (2.02% World share) | 18792 (2.51%) |
| Nr of word occurrences | 144.597 | 177.707 |
| Included in the analysis | 105 words which occur more than 160 times | 102 words which occur ≥ 190 times |
| Included with cosine ≥ 0.1 (pictures) | 68 words | 49 words |

Although South Korea and the Netherlands are comparable in terms of their number of scientific publications and of title word occurrences, the sets for the two countries produce very different looks when we apply social network visualization techniques to the semantic analysis of title words. This is illustrated in the Figures 4 and 5 by providing the network representation of the top hundred words in both sets. For the reader's information, the nodes represent words and the thickness of a line is proportional to the cosine values between the words distributions as vectors. Although both sets are not strongly organized—because the sciences are not primarily integrated in terms of nation states—the pictures show the different foci in the research portfolio of these two nations. The Korean set is organized in terms of the natural science disciplines with one additional cluster representing 'Asian medicine.' In other words, natural science fields such as materials, electronic control, and organic chemistry are very visible. This indicates that Korean academicians are traditional in their publication behaviour. While South Korea is weak in bio-medical sector, the Dutch set is highly focused on biomedicine and biotech. This focus accords with the one of the ISI database. About 80 percent of ISI journals are in the bio-medical field.

[7] The total number of records in the CD-Rom version of the *SCI 2002* is 784,458.

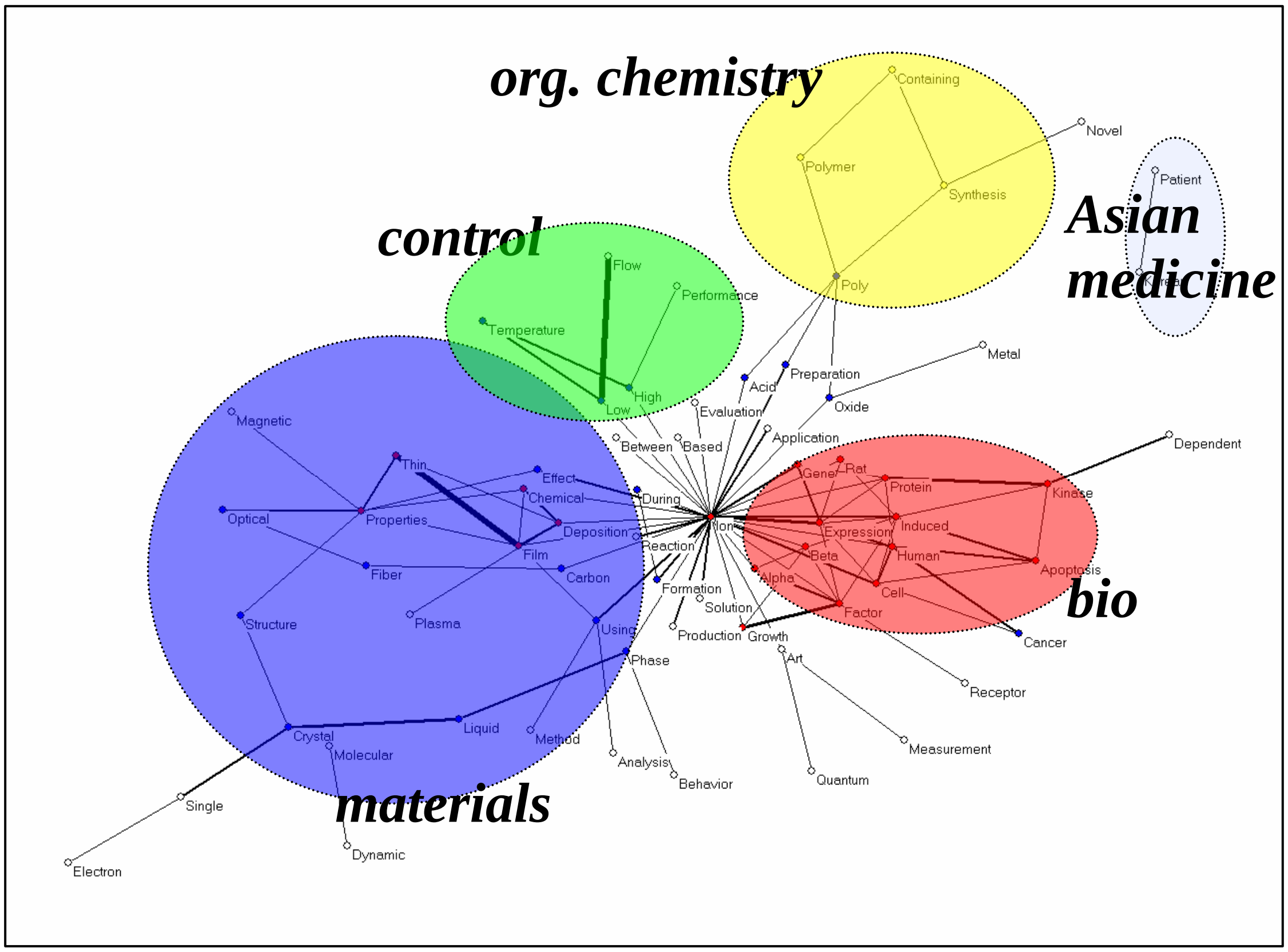

**Figure 4:** South-Korean set of publications covered by the *Science Citation Index 2002*; 68 most frequently used words with cosine ≥ 0.1.

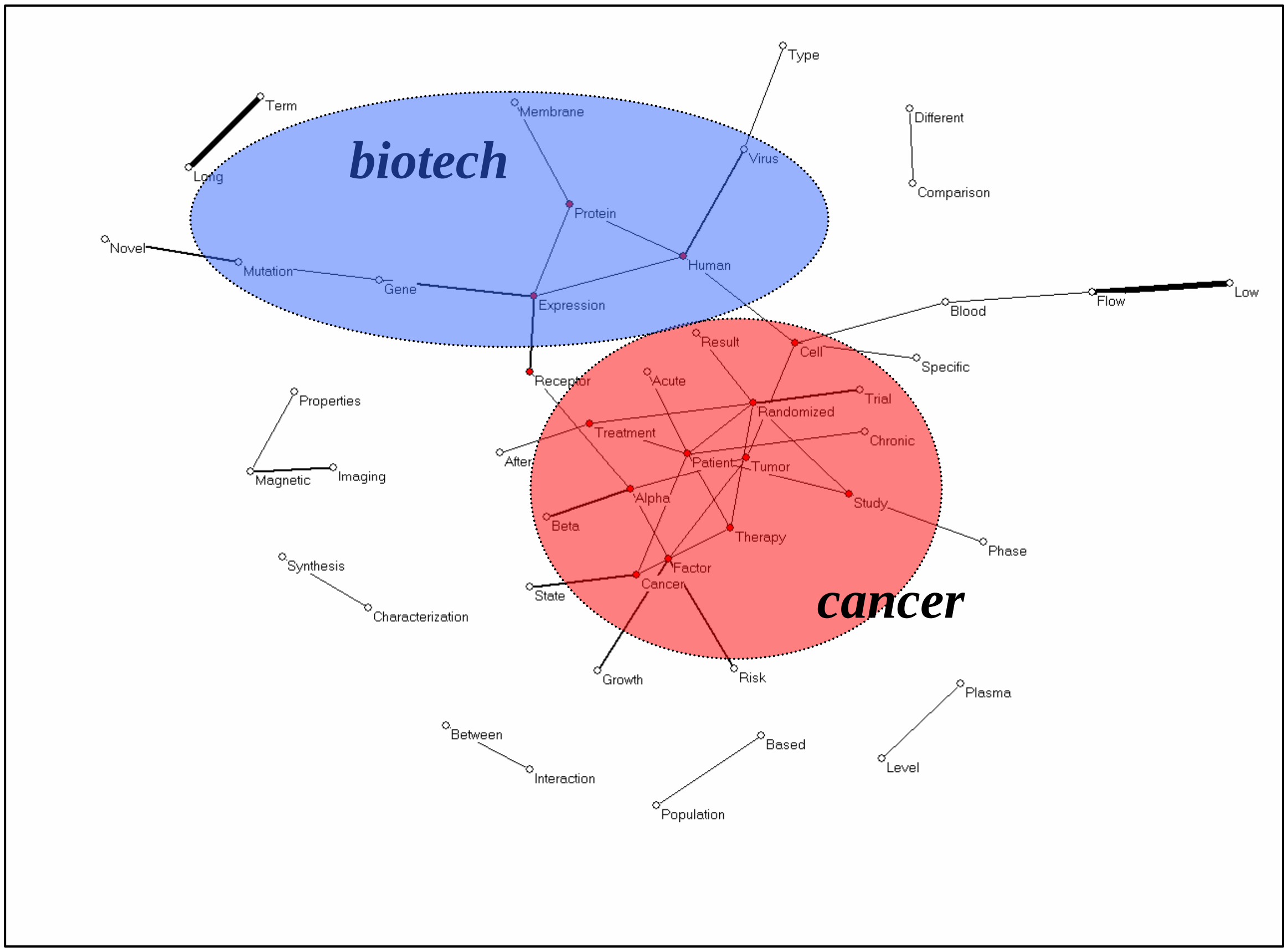

**Figure 5**: Dutch set of publications covered by the *Science Citation Index 2002*; 49 most frequently used words with cosine ≥ 0.1 for the Netherlands

## c. Technology Indicators

Before moving onto this section, it needs to be noted that one of us compared the Dutch portfolio with the university-based one in another context (Leydesdorff, 2004). Therefore, we do not present detailed results about Dutch patents in this research while the South Korean patents are relatively analyzed in depth. However, we will make comparisons between South Korea and the Netherlands wherever useful for extending the understanding of the differences and similarities between them.

*Data-gathering methods*

Historically, patent data bases have provided us with the oldest indicators. The U.S. Patent and Trademark Office (USPTO) makes all its patents available on-line at http://www.uspto.gov/ with images for the period 1790-1975, but full-text searchable since 1975. The World Patent database can be researched from the website of the European Patent Office at http://ep.espacenet.com/. Unlike national patent databases, the U.S. patents indicate an investment in the global marketplace (Grupp & Schmoch, 1999; Narin & Olivastro, 1992). Nowadays, these investments are made because of a value of the intellectual property to be internationally protected.

A patent contains a wealth of information that can be used to reveal a knowledge-based economy. For example, patents are linked to the scientific literature by citations. The so-called "non-patent literature references" (NPLR) contain references to scientific journal literature and book chapters among other things. Abbreviations of journal names, however, are not standardized. In the case of scientific references, most patents provide titles between quotation marks in order to distinguish them from journal names or from the title of an edited volume. We will use this indicator as a point of access for exploring the knowledge base of patents. Because the practice of using quotation marks is almost exclusively the case for formalized literature, we hypothesize that this indicator can be used as a proxy for accessing the knowledge base of patents.

*Results*

In 2002, there were 4,200 Korean patents in the USPTO database with at least one Korean address among inventors and/or assignees. It happens to be the case that there are 1,963 patents with a Dutch assignee and equally 1,963 patents with a Dutch inventor. The combined set, however, contains 2,827 patents with a Dutch address (2,824 of these patents could be retrieved). Table 6 provides the descriptive statistics for the datasets of the two countries, respectively.

The comparison between the results of the Netherlands and South Korea indicates that there are several factors evident. First, there is a large overlap between inventors and assignees in the Korean case. In the Dutch set we found both more co-inventors and co-assignees. The number of patent references is of the same order of magnitude in the two sets. Unique words in NPLR are smaller in the Korean set. The pattern of South Korea seems much more high-tech than that of the Netherlands but much less knowledge-based (in terms of NPLR) than the latter.

**Table 6:** Comparisons of the number of patents in the USPTO

| *2002* | *Netherlands* | *South Korea* |
|---|---|---|
| Nr of patents in USPTO[8] | 2,824 | 4,200 |
| Nr of assignees | 6,815 | 4,066 |
| Assignees based in NL and KR, resp. | 1,963 | 3,744 |
| Nr of inventors | 16,405 | 9,413 |
| Inventors based in NL and KR, resp. | 1,963 | 4,100 |
| Nr of unique title words | 4,005 | 3,984 |
| Nr of references to other patents | 31,514 | 36,972 |
| NPLR | 6,396 | 3,814 |
| NPLR with "" | 3,440 | 2,047 |
| Unique words in NPLR | 6,072 | 3,980 |
| Nr of patents with NPLR and "" | 643 | 440 |

Figure 6 provides a graphical representation of the intellectual structure of South Korea's patent portfolio in the USPTO. The 4,200 Korean patents contain 3,984 (stopwords excluded)[9] which occur 28,890 times. One hundred three words occur more than 40 times and are used for this analysis. Ninety six of these 103 words are connected to one another at cosine ≥ 0.1. The other seven words were dropped. Figure 6 exhibits the dominance of information and communication technologies (ICTs) and their applications. The term 'semi-conductor' and its neighbor words form the largest cluster near the origin of the map. This cluster gets connected to the words 'memory', 'circuit', and 'integrated'. Semi-conductors are technologically essential for the manufacture and operation of ICT. For the last few decades, South Korea's chip exports to the U.S.A. and other countries have been boosting the national economy as well as the Korean innovation system.

[8] The precise queries were as follows: "isd/$/$/2002 and (acn/kr or icn/kr)" and "isd/$/$/2002 and (acn/nl or icn/nl)", respectively. The abbreviations "kr" and "nl" are used for Korea and The Netherlands; "acn" is the field code of the name of the country of the assignee and "icn" for the name of the country of the inventor.

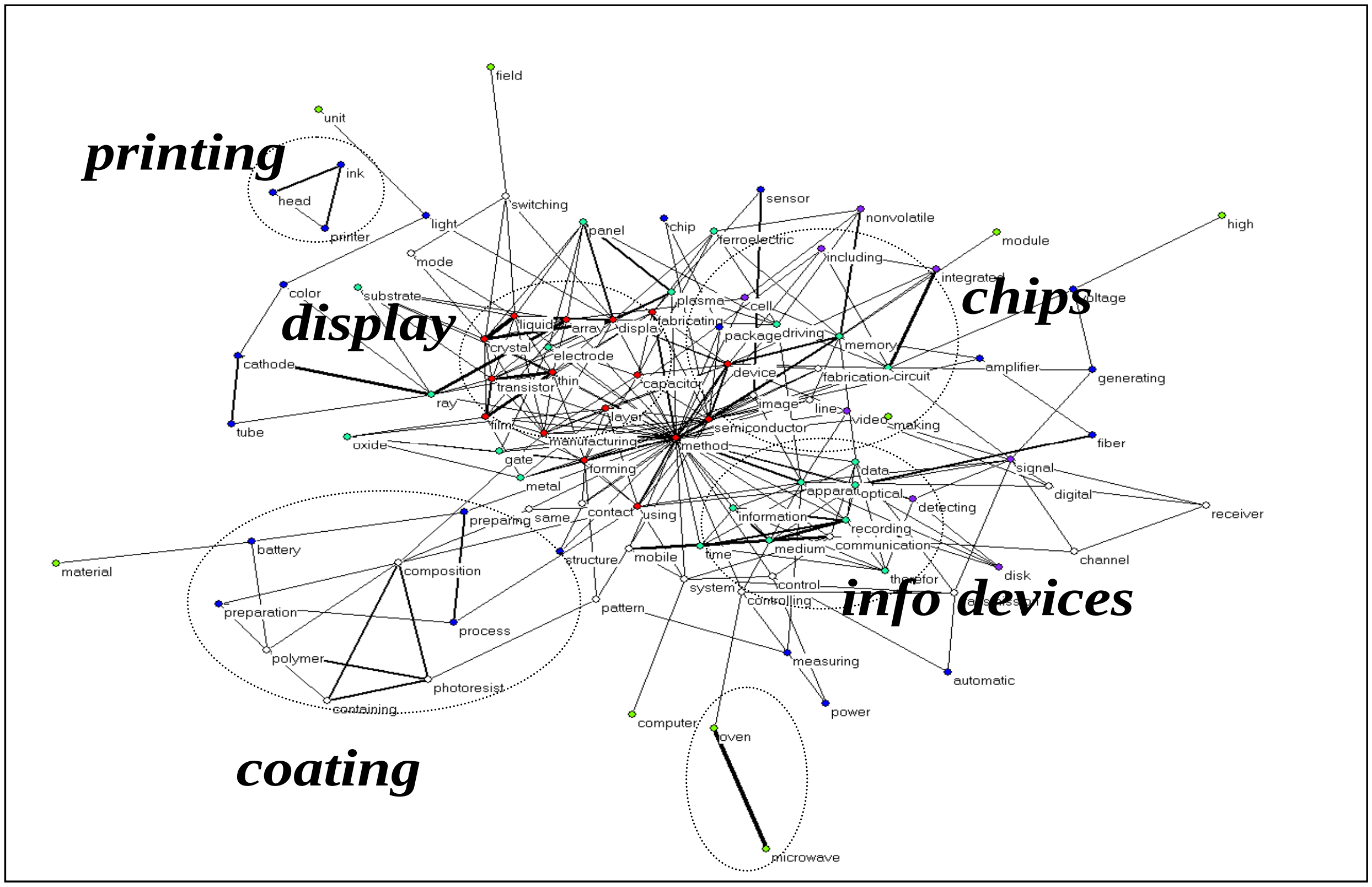

**Figure 6**: Cosine normalized map of 103 co-occurring words in patents (2002) with a Korean address among the assignees or inventors (Number of patents is 4,200; Word frequency > 40; 96 words connected at the threshold level of cosine ≥ 0.1).

There is a relatively tight grouping of 'liquid', 'crystal', and 'display' exhibiting that South Korea is globally leading in the display industry. The next most prominent industries of South Korea (fiber optics business, mobile medium, and communication device) are found around the central group. The words such as 'mobile', 'medium', and 'communication' are assorted west to east, from the left to the right. Lastly, the map clearly partitions relatively peripheral industries (e.g., electro-technical and chemical applications) from central ones in South Korea. Industrial technologies are interspersed, making a circle centered on the word 'semi-conductor'. On the left side, a group of words ranging from 'polymer' to 'composition' form a triangle differentiating themselves from post-industrial ICTs. In the upper-left corner, a small set of patents related to a printer industry clearly illustrate a distinctive cluster of computer hardware.

[9] The stopword list of the USPTO database (at http://www.uspto.gov/patft/help/stopword.htm) was used throughout this study.

In summary, the results of cosine map show that the words related to digital technologies are placed closely together in a center indicating that South Korea is a global supplier of integrated chips, semi-conductor, computer peripherals, and information devices. The less-digitalized (or more industrial technologies) are scattered on the side. One of interesting findings is that we can't see a sign of steel (or iron industry) from the patent map of South Korea. This implies that shipping industry used to be very strong in the past but it is declining.

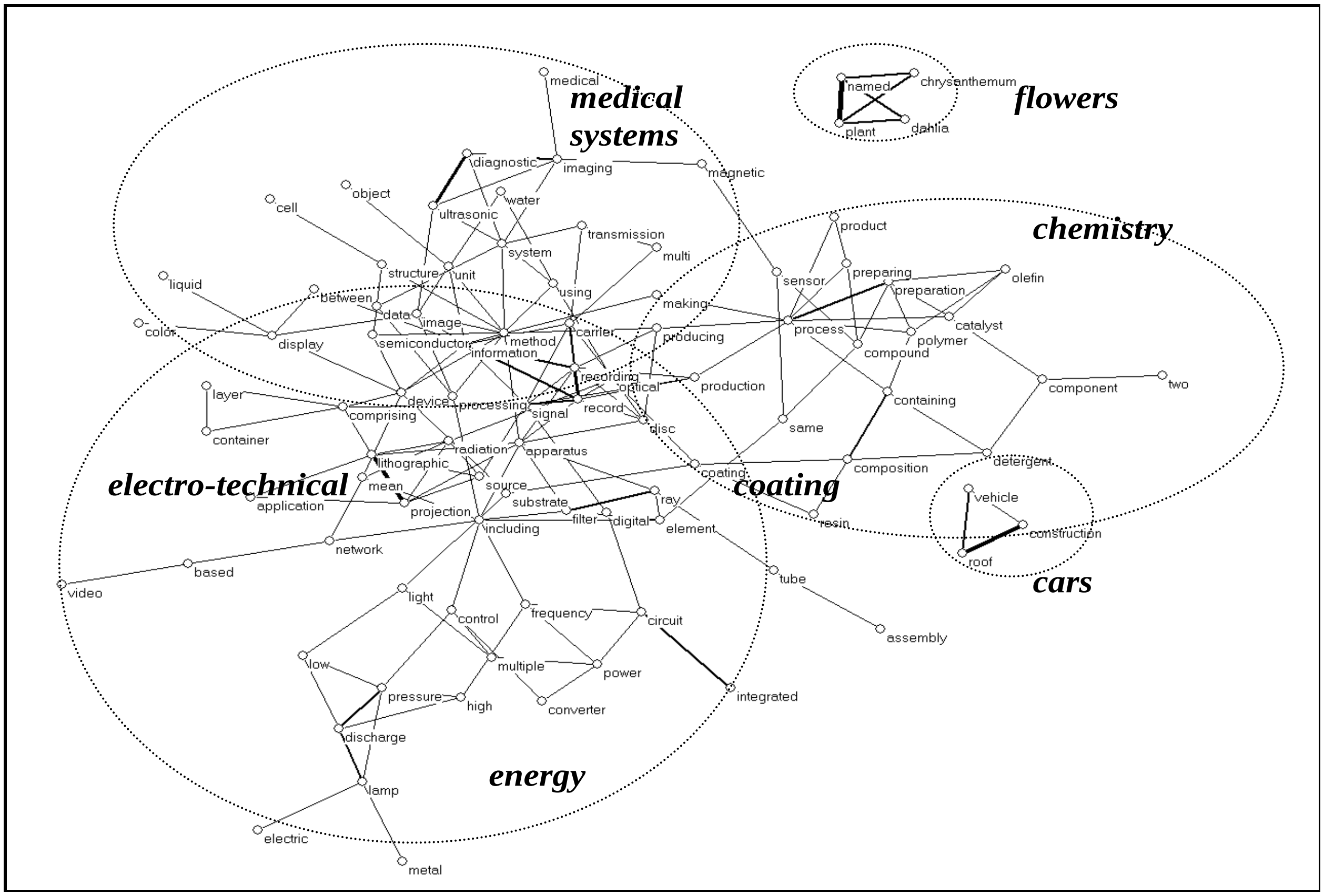

**Figure 7:** Cosine normalized map of 105 co-occurring words in patents (in 2002) with a Dutch address among the assignees or inventors (N Patents = 2,824; Word frequency > 22; cosine ≥ 0.1).

The corresponding visualization for the Netherlands (Figure 7) exhibits a recognizable representation of the Dutch industrial structure with a dominance of electro-technical and chemical applications. Multinational corporations are dominant in the set. For example, Philips with a focus on electro-technical systems holds 768 of the 1,963 patents (39.1%) with a Dutch address among the assignees. Medical systems are related to the electro-technical side of the set through imaging devices. The occurrence of a small set of patents related to the names of flowers is noteworthy.

Figure 8 shows the network of title words in patents with a Korean address in relation to the title words used in their literature references. The white dots represent the patents, the black ones the non-patent literature references (scientific literature) cited by the patents. The corresponding picture for the Netherlands (Figure 9) shows a clear group of biomedical patents with a halo of title words from NPLR, but in the Korean case this position is occupied by words which indicate information technologies. Patents in this sector are science-based, but as we shall see in the next section the development of these sciences has remained underdeveloped in terms of publications with a Korean address. Perhaps, the knowledge-base of these patents is imported at the global level.

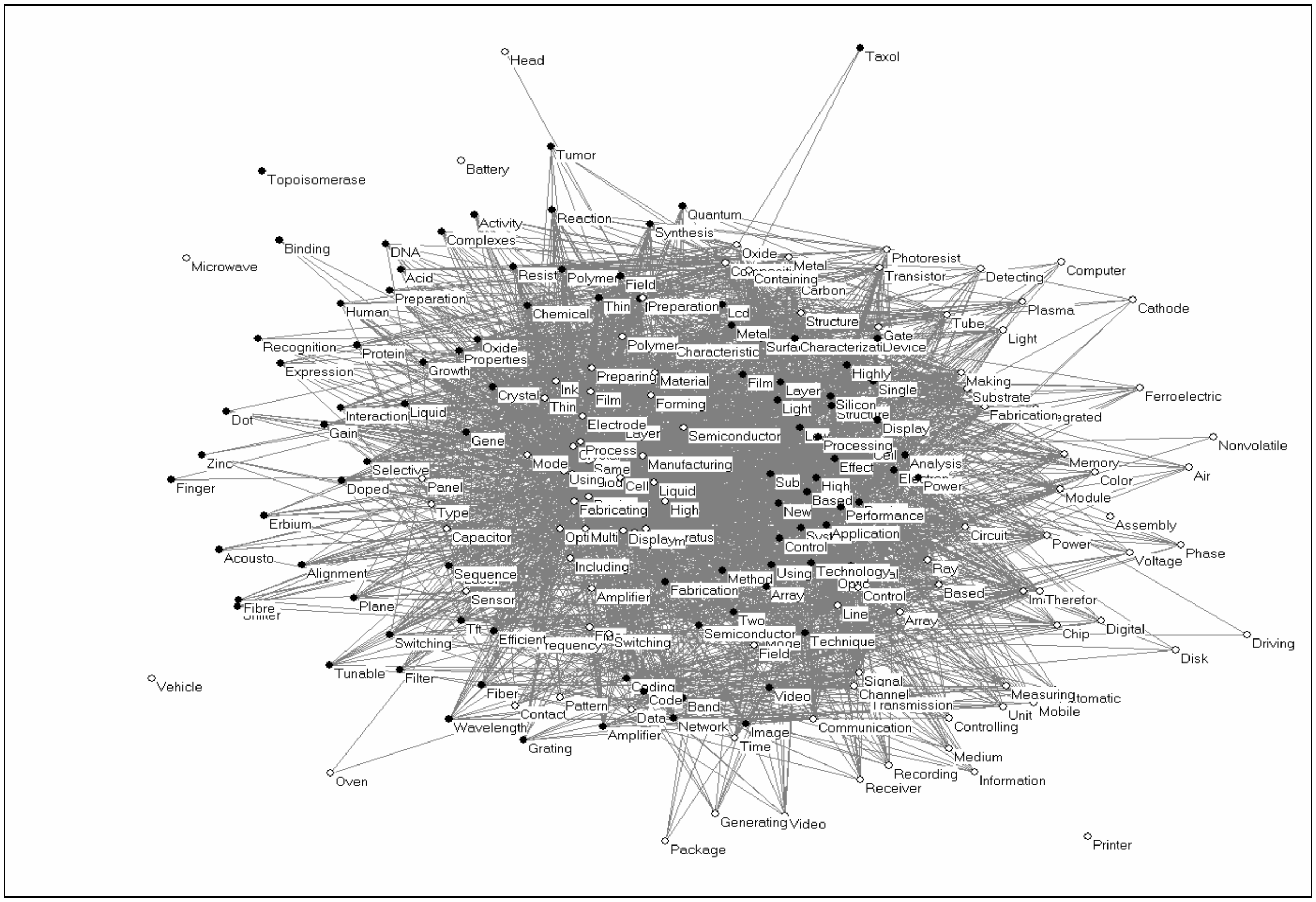

**Figure 8:** Network of title words in patents with a Korean address among the assignees or inventors in relation to the title words used in their literature references (N Patents = 4,200; Word frequency > 40; 2,047 literature references with 3,908 unique words of which 101 occur with a frequency > 24).

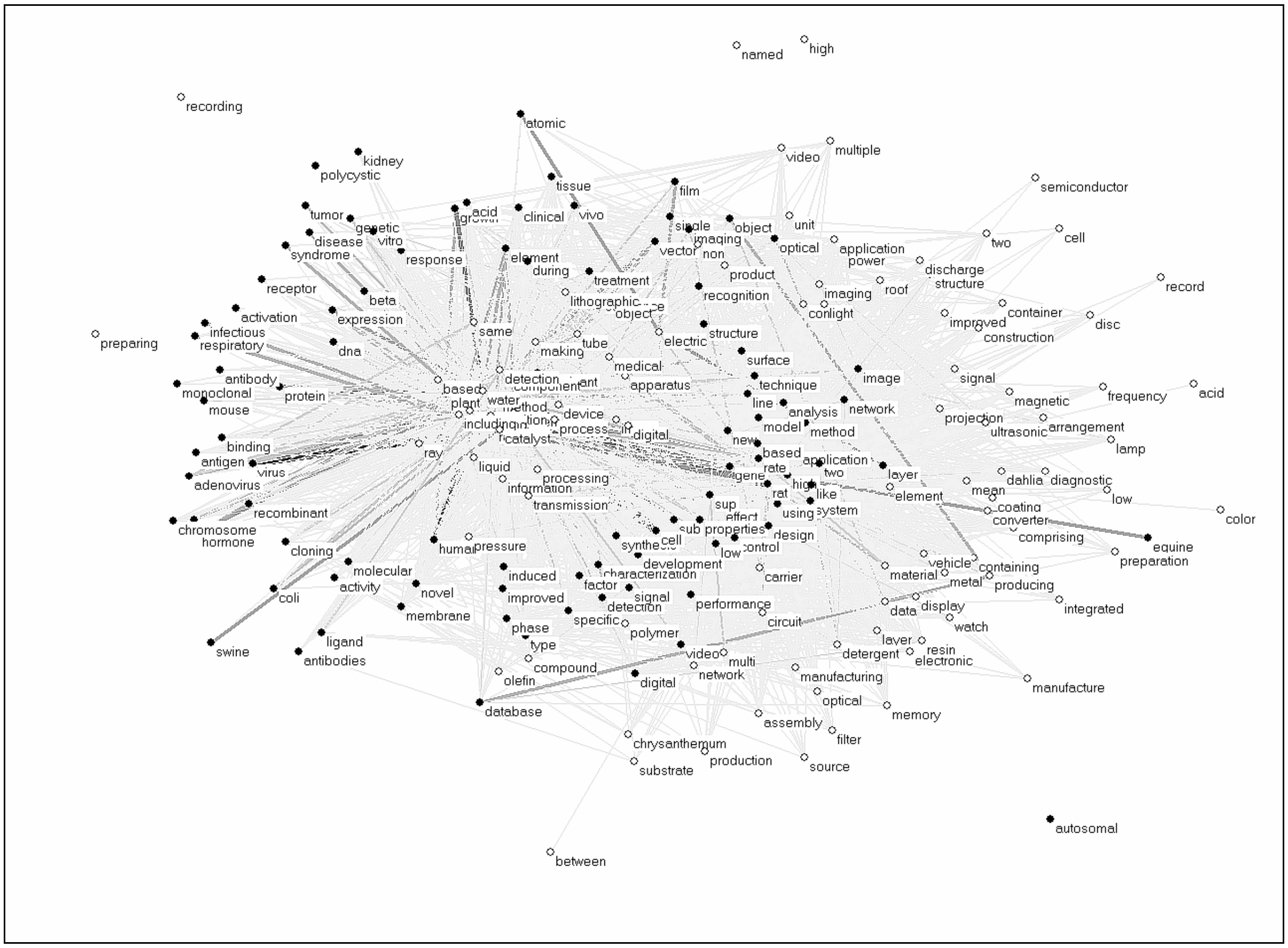

**Figure 9:** Network of title words in patents with a Dutch address among the assignees or inventors in relation to the title words used in their literature references (N Patents = 2,824; Word frequency > 22; 3,440 literature references with 6,072 unique words of which 101 occur with a frequency > 31). (Source: Leydesdorff, 2004, at p. 998.)

Figure 10 shows that the knowledge base of South-Korean patents is focused on ICT applications (e.g., information devices) more than in other patent sets. Compared to Figure 6, however, the patents with respects to the semi-conductor are no longer central to the aggregate of the ICTs in Figure 10. While the terms of display industry (e.g., liquid, crystal, and display) are related to the semi-conductor side in Figure 10, they are now forming a thick square on the lower right corner reflecting the intensive knowledge inflows from a scientific sector to the industry. A second cluster is focused on communication technologies like reproduction and connection.

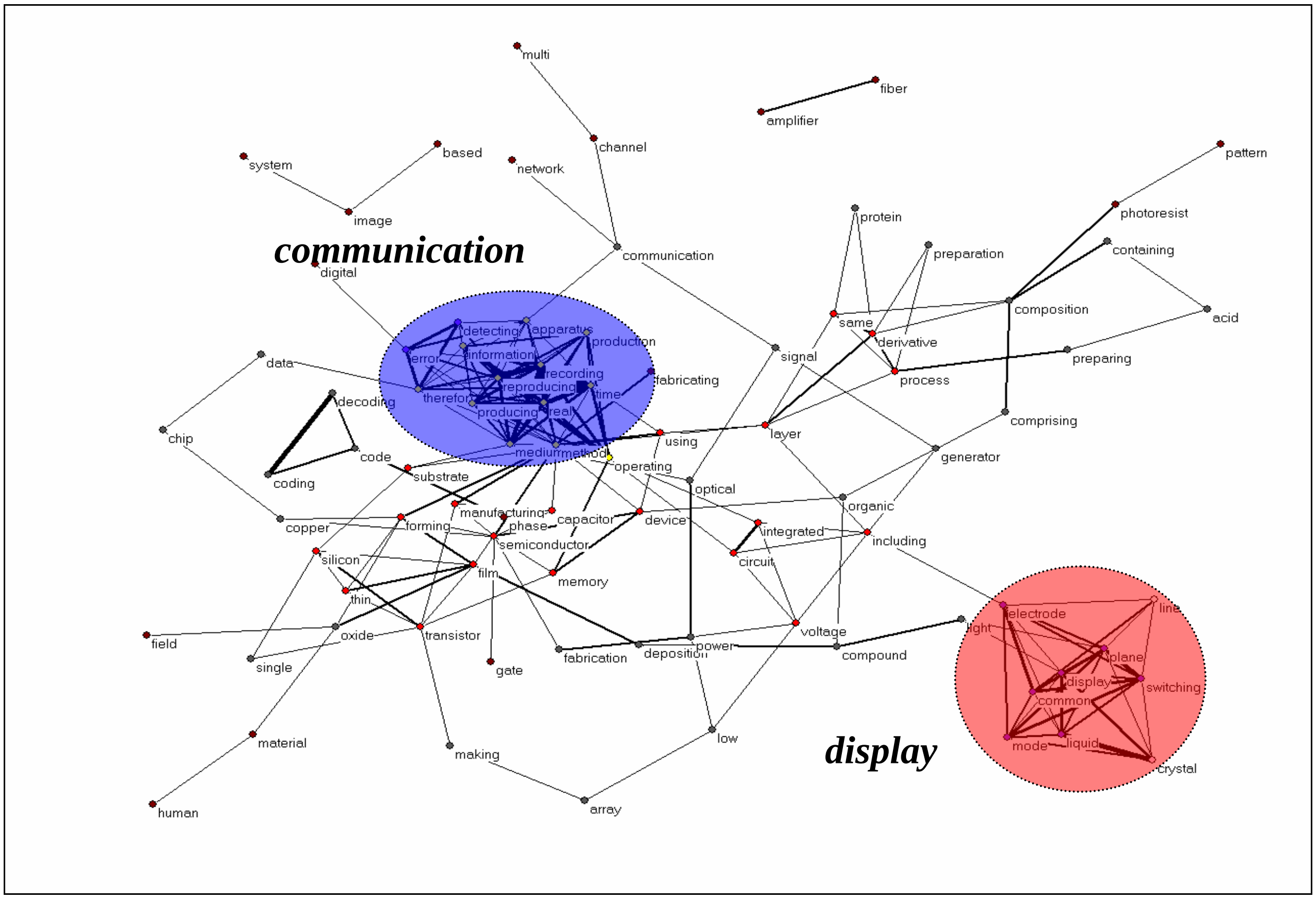

**Figure 10:** Cosine normalize map of 95 most frequently occurring words in 440 "literature-based" patents with a Korean address among the assignees or inventors (N Patents = 440; Word frequency > 5; 88 words connected at the threshold level of cosine ≥ 0.2)

In the case of the Netherlands, the biomedical applications are not so visible in the map of "science-based" patents Figure 11 as in Figure 9. Figure 11 shows that the industrial structure remains more important than the intellectual organization of biomedical patents. Biomedical terms (e.g., "DNA," "nucleic") are relatively peripheral in Figure 11. However, the finding that the knowledge base of this patent set is integrated by a biomedical network of title words in their NPLR is meaningful because the industrial structure visible at the surface is dominated by electro-technical and chemical applications.

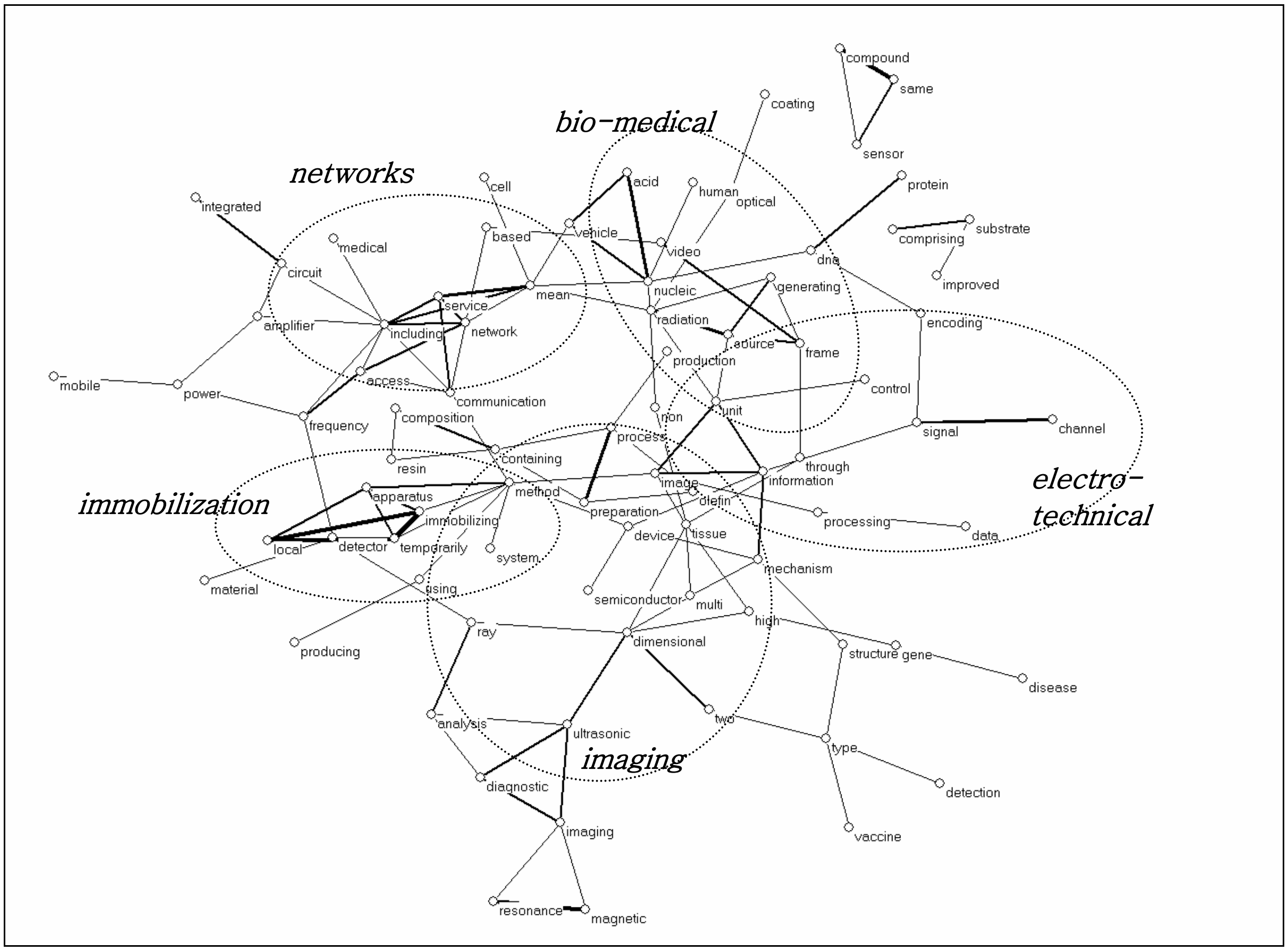

**Figure 11**: Cosine normalized map of 107 most frequently occurring words in 643 "literature-based" patents with a Dutch address among the assignees or inventors (N Patents = 643; Word frequency > 6; 83 words connected at the threshold level of cosine ≥ 0.2). (Source: Leydesdorff, 2004, at p. 999.)

## CONCLUSION

In this paper we made an attempt to compare the knowledge bases of South Korea and the Netherlands in various dimensions of Triple Helix mode of relations. The comparison was made using webometric, scientometric, and technometric indicators for 2002. We could clearly retrieve the differences between these two national systems of innovation. The conclusion was the South Korea has a strong patent portfolio in the USPTO database, but that a relation with the knowledge base of this portfolio is not visible in terms of publication patterns of Korean scholars. The publication patterns of Dutch scholars, on the other hand, demonstrate a clear orientation towards biomedical research and biotechnology, but this is hardly visible in the patent portfolio of the Netherlands in the USPTO database. This portfolio is completely dominated by existing industrial structures. In the Korean case, however, the traditional industrial structures like ship-building and steel are not visible in the patent portfolio.

A second purpose of this study was to demonstrate the use of three-dimensional transmissions as a methodology for data analysis. Few papers have offered an indicator for the comprehensive analysis of cross-national innovation system (cf. Nelson, 1993). Data collection may require more care. However, independently of the refinement of the measurement, network data about university-industry-government relations can usually be written as relative frequency distributions. The indicators of the three-dimensional transmissions can then be applied to a comparison of the state of the Triple Helix configurations under study.

It would be interesting to extend this macro-data to the year 2004 and to follow up with more detailed and precise questions and discussions. Time series can also be tested on the emergence of new systemness (Leydesdorff, 1995; Leydesdorff & Scharnhorst, 2002). Patent data can also be analyzed in terms of the distribution over industrial sectors (patent classification categories) or in comparison with competing countries in the respective regions. For example, the two nations, South Korea and the Netherlands can also be compared in terms of their relative position in comparison with major economic systems in their geographical environments.

From the perspective of the further development of webometrics, the types of webpages and the information contained on these pages can be classified based on the categories such as secondary national domains (e.g., webpages of South Korean academic organizations end with .ac.kr) or the taxonomy schemes of the search engine at Yahoo.com. The knowledge and information bases of social systems can also be compared in different dimensions. The comparison of national systems of (post-industrial) knowledge, (science and technology-based) innovations, and (digital) information provides an agenda for future research.

This research has the following surplus values for the policy programs in both South Korea and the Netherlands. The primary implications of this research reside in examining the configurations of national knowledge-based systems inscribed in science, technology, and innovation networks using the Triple Helix indicators. Despite the increasing amount of scientific and technological outputs in terms of the knowledge-based dynamics, South Korea's portfolio is more traditional than that of the Netherlands in both public and private sector. For example, research and patenting in the biomedical sector is underdeveloped. In terms of the Internet-economy, the Netherlands seems to have more service-oriented components than South Korea does. Thus, this research provides a starting point for cross-country evaluation of national innovative policies influencing knowledge networks. For South Korea, this international comparison suggests local policy-makers to facilitate the expansion of scientific and innovative research to biomedical area and biotechnology bringing together relevant academic and professional communities. With respect to the Netherlands, the primary finding of this study is perhaps useful for the design of a certain type of research policy. Given the dominant role of Dutch universities at a network level, this identification may help the authorities concerned to take a policy action in order to address the challenges of inter-personal, inter-institutional, and/or inter-disciplinary collaboration in more precise terms. Dutch policy-makers may derive specific targets to further develop university-industry cooperation. In general, the results of this research provide policy-makers some insight into the determinants and consequences of (in)formal interaction between public and private research sector since Triple Helix indicators and their accompanying semantic mappings produce a networked topology of knowledge-based innovation system in both South Korea and the Netherlands.